\documentstyle[aps,multicol,epsfig]{revtex}

\draft
\begin{document}

\title{{\bf Complete quantum teleportation with a Kerr nonlinearity}}
\author{David Vitali, Mauro Fortunato, and Paolo Tombesi}
\address{Dip. di Matematica e Fisica and Unit\`a INFM, Universit\`a di
Camerino, via Madonna delle Carceri 62032, Camerino, Italy}

\date{\today}

\maketitle

\begin{abstract}
We present a scheme for the quantum teleportation of the polarization
state of a photon employing a cross-Kerr medium. 
The experimental feasibility of the scheme is discussed and 
we show that, using the recently demonstrated ultraslow light propagation
in cold atomic media, our proposal can be realized with presently
available technology.
\end{abstract}



\begin{multicols}{2}

Quantum entanglement is a powerful resource at the basis of the extraordinary
development of quantum information.
Among the most fascinating examples of the possibilities offered by 
sharing quantum entanglement are quantum teleportation~\cite{qt93},
quantum dense coding~\cite{benwies}, entanglement 
swapping~\cite{swap}, quantum cryptography~\cite{crypt}, and quantum 
computation~\cite{qucomp}.
Quantum teleportation is the ``reconstruction'', with $100\%$ success,
of an unknown state
given to one station (Alice), performed at another remote station (Bob),
on the basis of two bits of classical information sent by Alice to Bob.
Perfect teleportation is possible only if the two parties share a maximally
entangled state. The most delicate part
needed
for the effective realization
of teleportation is the Bell-state measurement, {\it i.e.} the discrimination
between the four, maximally entangled, Bell states~%
\cite{bell}
which has to be performed by Alice and whose result is communicated to Bob
through the classical channel.
There have been numerous proposals for its
realization in different systems~\cite{dav} and recently
successful, pioneering experiments~\cite{zei,dem,kim} have provided
convincing experimental proof-of-principle of the correctness of the
teleportation concept.

These experiments differ by the degrees of freedom
used as qubits and for the different ways in which the Bell-state
measurement is performed. The Innsbruck experiment~\cite{zei} is
the conceptually simplest one, since each qubit is represented by
the polarization state of a single photon pulse. In this experiment, 
however,
only two out of the four
Bell states can be discriminated and therefore
the success rate cannot be larger than $50 \%$ \cite{brau}.
The Rome experiment~\cite{dem} employs the entanglement between the spatial
and the polarization degrees of freedom
of a photon and it is able to
distinguish all the corresponding four Bell states completely.
However in this scheme the state to be teleported 
is generated within the apparatus
(it cannot come from the outside)
and therefore the scheme cannot be used as a computational primitive
in a larger quantum network for further information processing, as it
has been recently proposed in Ref.~\cite{gotte}.
Finally the Caltech experiment~\cite{kim} is conceptually 
completely different since it implies the teleportation
of the state of a continuous degree of freedom~\cite{contin},
the
mode
of an electromagnetic field, employing the entangled
two-mode squeezed states
at the output
of a parametric amplifier.
In this case, the Bell-state 
measurement is replaced by two homodyne measurements and a
direct comparison with the original quantum teleportation scheme of
Ref.~\cite{qt93} cannot be made.
Up to now, only coherent states of the electromagnetic field have 
been successfully teleported using this scheme.

It is therefore desirable to have a scheme for a Bell-state measurement
that can be used in the
simplest
case of the Innsbruck scheme. This would
imply the possibility
of realizing
the first {\em complete} verification
of the original quantum teleportation scheme \cite{qt93} and also
of having a device useful for other quantum protocols, as quantum dense coding
\cite{benwies}.
What we need is a device able to
discriminate among
the four Bell states
that can be realized with the polarization-entangled photon pairs
produced in Type-II phase matched parametric down conversion \cite{kwiat},
that is
\begin{mathletters}
\label{bell}
\begin{eqnarray}
\label{bellst1}
|\psi^{\pm} \rangle & = & \frac{|V_1,H_2\rangle \pm |H_1,V_2 \rangle}{\sqrt{2}}
 \\
|\phi^{\pm}\rangle & = & \frac{|V_1,V_2\rangle \pm |H_1,H_2 \rangle}{\sqrt{2}}
  \;,
\label{bellst}
\end{eqnarray}
\end{mathletters}
where $|H\rangle $ and $|V\rangle $
denote the horizontally and vertically polarized one-photon states,
respectively, and $1,2$ refer to two different spatial modes.

It has been recently shown that it is impossible to perform a 
complete Bell measurement on two-mode polarization states using only
linear passive elements \cite{lutk} (unless the two photons are
entangled in more than one degree of freedom \cite{kwwe}), 
and for this reason schemes
involving some effective nonlinearities, such as resonant atomic interactions
\cite{scul}, or the Kerr effect \cite{paris}, have been proposed.
In the present Letter we propose a
scheme for a perfect Bell-state discrimination based on a nonlinear
optical effect, the cross-phase modulation 
taking place in Kerr media.
In this respect, our scheme is based on a 
$\chi^{(3)}$ medium as the ``Fock-filter'' proposal of Ref.~\cite{paris}.
However, our scheme is different and simpler and, above all, is feasible using
available technology, since we shall show that the needed crossed-Kerr
nonlinearity can be obtained using the recently demonstrated ultraslow
light propagation \cite{hau},
achieved via electromagnetic induced transparency (EIT) \cite{eit} in
ensembles of
cold atoms. 

\begin{figure}[bt]
\centerline{\epsfig{figure=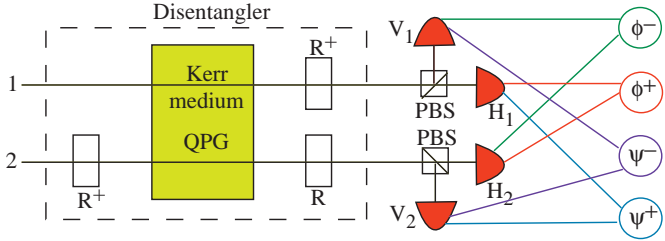,height=4cm}}
\narrowtext
\caption{Scheme of the Bell-state measurement: QPG is the quantum 
phase gate of Eqs.~(\protect\ref{qpg}) with
$\varphi=\pi$; $R$ ($R^{\dagger}$) rotates the polarization by 
$\pi/4$ ($-\pi/4$), and PBS are
polarizing beam splitters. The 
``disentangler'' performs the unitary transformation of
Eqs.~(\protect\ref{belltra}).}
\label{fig1}
\end{figure}

Our ``Bell box'' is described in Fig.~1 and
can be divided into two
parts: the left part is composed by three polarization rotators
($R$, $R^{\dagger}$) and by the ``quantum phase gate'' 
(QPG) which will be described 
below, and can be called ``the disentangler'', since it 
realizes the unitary transformation 
changing each Bell state of Eqs.~(\ref{bell})
into one of the four factorized polarization states, {\it i.e.}
\begin{mathletters}
\label{belltra}
\begin{eqnarray}
\label{belltra1}
|\psi^{+}\rangle &\rightarrow & 
|H_1,V_2\rangle \\
|\psi^{-}\rangle &\rightarrow & 
|V_1,V_2\rangle \label{belltra2} \\
|\phi^{+}\rangle &\rightarrow & 
|H_1,H_2\rangle \label{belltra3} \\
|\phi^{-}\rangle &\rightarrow & 
|V_1,H_2\rangle \;.
\label{belltra4}
\end{eqnarray}
\end{mathletters}
The right part of the scheme is composed by two
polarizing beam-splitters
(PBSs) and by four detectors with single-photon
sensitivity, and simply serves the purpose of detecting
the four states of the factorized polarization basis 
\begin{eqnarray}
&&\left\{|e_{1}\rangle,
|e_{2}\rangle,|e_{3}\rangle,|e_{4}\rangle\right\} \label{basis} \\
&&= \left\{|H_1,H_2\rangle,
|H_1,V_2\rangle,|V_1,H_2\rangle,|V_1,V_2\rangle\right\} \nonumber\;,
\end{eqnarray}
where $\{|e_{i}\rangle\}$ are the tensor product of the single-photon
polarization basis states, 
\begin{equation}
|H_{i}\rangle = \left(\matrix{
		1 \cr
		0 \cr}\right)_{i} \;\;\;\;\;\;\;\;
		|V_{i}\rangle =
		\left(\matrix{0 \cr 1 \cr}\right)_{i}.
\label{base}
\end{equation}
Due to the one-to-one correspondence of 
Eqs.~(\ref{belltra}),
it is clear that the detection of each Bell state 
corresponds to a different pair of detector clicks, so that they are
unambiguously distinguishable. 
The disentangler, and in particular the QPG, is the 
most delicate part as concerns the experimental implementation, 
since it involves a two-qubit operation, {\it i.e.}, an effective 
photon-photon interaction.
In fact, if $\tilde{R}_{i}$ is a simple polarization rotation by
$\pi/4$ radians for mode $i$ (and $\tilde{R}_{i}^{\dagger}$ its inverse),
{\it i.e.},
$|H_i\rangle \rightarrow 
\left(|H_i\rangle + |V_i\rangle \right)/\sqrt{2}$,
$|V_i\rangle \rightarrow 
\left(|V_i\rangle - |H_i\rangle \right)/\sqrt{2}$, we have
\begin{equation}
R_{1}=\tilde{R}_{1}\otimes I_{2}\;, R_{2}=I_{1}\otimes 
\tilde{R}_{2}\;,
\label{pola}
\end{equation}
which can be obtained using a $\lambda/2$ retardation plate at a $\pi/8$ angle.
In Eq.~(\ref{pola}) $I_{i}$ is the $2 \times 2$ unit matrix for mode 
$i$.
The general QPG $P(\varphi)$ is a universal two-qubit gate
as long as $\varphi \neq 0$~\cite{qpg,turch}, and in the two-photon
polarization basis~(\ref{basis})
we are considering here, it can be written as
\begin{mathletters}
\label{qpg}
\begin{eqnarray}
|H_1,H_2\rangle &\rightarrow & 
|H_1,H_2\rangle \label{qpg1} \\
|H_1,V_2\rangle &\rightarrow & 
|H_1,V_2\rangle \label{qpg2} \\
|V_1,H_2\rangle &\rightarrow & 
|V_1,H_2\rangle \label{qpg3} \\
|V_1,V_2\rangle &\rightarrow & 
e^{i\varphi}|V_1,V_2\rangle \;.
\label{qpg4}
\end{eqnarray}
\end{mathletters}
The experimental realization of this gate has been reported in 
Ref.~\cite{nist}, in the case 
when one qubit is given by
the internal state of a trapped ion
and the other qubit by its two lowest vibrational states, and recently
in Ref.~\cite{ens}, where the two qubits are represented by two circular
Rydberg states of a Rb atom and by the two lowest Fock states of a microwave 
cavity. In the optical case we are interested in, the QPG between
two frequency-distinct cavity modes has been
experimentally investigated in Ref.~\cite{turch}, using however weak
coherent states instead of single photon pulses, demonstrating therefore
only conditional quantum dynamics and not the full quantum transformation
of Eqs.~(\ref{qpg}). As it can be easily checked,
the QPG~(\ref{qpg})
can be realized using a crossed-Kerr
interaction involving the vertically polarized modes only
\begin{equation}
H_K = \hbar \chi a_{V_1}^{\dagger}a_{V_1}a_{V_2}^{\dagger}a_{V_2} \;,
\label{croke}
\end{equation}
so that the conditional phase shift is $\varphi = \chi t_{int}$, where
$t_{int}$ is the interaction time within the Kerr medium.

The disentangler of Fig.~1 realizes the 
transformation (\ref{belltra}) 
when the conditional phase shift is
$\varphi = \pi$, as it can be checked in a straightforward way
by writing the matrix form of the transformation
\begin{equation}
R^{\dagger}_{1}R_{2}P(\pi)R^{\dagger}_{2}
\label{transf}
\end{equation}
of Fig.~1 in the factorized polarization basis~(\ref{basis}),
which is just the matrix form of Eqs.~(\ref{belltra}) in the chosen basis.
The proposed Bell box is therefore extremely simple and also
robust against detector inefficiencies. This is due to the fact that
in our scheme, only one photon at most impinges on each of the 
four detectors. First of all this means that
only single photon {\em sensitivity} and not single photon
{\em resolution} is needed, and in this case solid-state 
photomultipliers can 
provide up to $90\%$ efficiency~\cite{kwwe}. Moreover, this implies that the
detection scheme is reliable, {\it i.e.}, it always discriminates the correct
Bell state, whenever it answers. In the case of detectors with the same
efficiency $\eta$, our Bell box gives the (always correct)
output with probability 
$\eta^{2}$ and it does not give any output (only zero or one photon is
detected) with probability $1-\eta^{2}$.

As we have already remarked, the most difficult part for the 
experimental implementation of the scheme is the QPG
with a conditional phase shift $\varphi = \pi$. In fact, realizing the
transformation~(\ref{qpg}) means having a large cross-phase 
modulation at the single photon level between two traveling-wave
pulses,
with negligible absorption, which is very demanding.
For example, in the experiment of Ref.~\cite{turch}, a conditional 
phase-shift $\varphi = 16^{\circ} $ has been measured, which however 
involved two frequency-distinct {\em cavity} modes in a high-finesse 
cavity. However, the recent demonstration of ultraslow light 
propagation in a cold gas of sodium atoms~\cite{hau} and with hot
Rb atoms~\cite{kash}, opens the way 
for the realization of significant conditional phase shifts also
between two traveling single photon pulses. In fact, the extremely 
slow group velocity is obtained as a consequence of EIT~\cite{eit}, 
which however, as originally 
suggested by Schmidt and Imamo\u{g}lu in Ref.~\cite{optlet}, can also 
be used to achieve giant crossed-Kerr nonlinearities. In fact, Harris 
and Hau~\cite{hhau}, developing the suggestions of Ref.~\cite{optlet},
showed that when the ultraslow group velocity
is the dominant feature 
of the problem, nonlinear optical processes between traveling
pulses with low number of photons  
become feasible.

In particular, in the limit of very small group 
velocity, and therefore with light pulses compressed to a spatial
length much smaller than the medium length, they find a 
conditional phase shift per photon between two pulses 
(characterized by frequencies
$\omega_{24}$ and $\omega_{p}$ in \cite{hhau}) given by 
$\varphi = \gamma_{24}\Delta \omega_{24}/
\left(4\gamma_{24}^{2}+4\Delta \omega_{24}^{2}\right)$, accompanied by
a two-photon absorption $\gamma_{24}^{2}/
\left(4\gamma_{24}^{2}+4\Delta \omega_{24}^{2}\right)$, where 
$\Delta \omega_{24}$ is the detuning of one of the two pulses and 
$\gamma_{24}$ the associated linewidth (see Eq.~(10) of Ref.~\cite{hhau}). 
For a sufficiently large detuning $\Delta \omega_{24} \gg \gamma_{24}$, 
two-photon absorption is negligible and we have just the desired result, 
{\it i.e.},
a significant conditional phase shift between two traveling single 
photon pulses without appreciable absorption. Unfortunately, in 
this same limit, the phase shift becomes $\varphi \simeq \gamma_{24}/
4\Delta \omega_{24}$ which cannot be too large and close to $\pi$, as we 
have assumed above in the Bell box scheme. This may be a problem
because it is possible to see 
that if the phase $\varphi$ of the QPG is not equal 
to $\pi$, the scheme of Fig.~1 is no
longer perfect and it does not 
discriminate the four Bell states with $100 \%$ success.
However, it should be noted that this is not a theoretical limitation,
but only a practical drawback of the specific scheme of 
Ref.~\cite{hhau}.
Furthermore, as mentioned above,
the QPG represented by $P(\varphi)$
is a universal two-qubit 
gate, capable of entangling and disentangling qubits as soon as 
$\varphi \neq 0$. Moreover, even though different from $\pi$, the 
conditional phase shift $\varphi$ is a given and measurable property,
and it is reasonable to expect that, using the knowledge of 
the actual value of $\varphi$, it is possible to adapt and optimize 
the teleportation protocol in order to achieve a truly {\em quantum}
teleportation ({\it i.e.}, that cannot be achieved with only classical
means), even in the presence of an imperfect Bell-state measurement.
Optimization means that Bob has to suitably modify
the four local unitary transformations he has to perform on
the received qubit according to the Bell measurement result 
communicated by Alice. In the optimized protocol, Bob's local 
unitary transformations will now depend on the phase $\varphi$ 
of the QPG and will reduce to those of the 
original proposal~\cite{qt93} in the ideal case of perfect
Bell-state discrimination $\varphi =\pi$. We expect that, $\forall 
\varphi \neq 0$, the average fidelity of the teleported state will be 
always larger than $2/3$, as it must be for any truly quantum 
teleportation of a qubit state~\cite{barnum}.

Let us therefore consider a generic one-photon state 
$|\psi \rangle_{1} = \alpha 
|H_{1}\rangle + \beta |V_{1}\rangle $, which is given to Alice and has 
to be teleported to Bob, and let us assume that Alice and Bob share the 
Bell state $|\psi^{+}\rangle_{23}=(|H_{2}V_{3}\rangle + 
|V_{2}H_{3}\rangle)/\sqrt{2}$, so that the input state for the 
teleportation process is $|\psi\rangle_{1} \otimes |\psi^{+}\rangle_{23}$.
Alice is provided with the ``imperfect'' Bell box with a
QPG $P(\varphi)$,
so that
the disentangler of Fig.~1 
will now be described by the
transformation $R^{\dagger}_{1}R_{2}P(\varphi)R^{\dagger}_{2}$.
It is easy to check that when $\varphi \neq \pi$, the four Bell states
are no
longer completely disentangled and therefore no
longer discriminated with $100 \%$ success.

Alice has to perform the Bell-state measurement on modes 1 and 2, and
the resulting
joint state of the 
three modes just before the photodetections is
\begin{equation}
	|\tilde{\psi} \rangle_{123}=
	\sum_{i=1}^{4}|e_{i}\rangle_{12} \hat{G}_{i}(\varphi)|\psi\rangle_{3}
	\label{befodete}\;,
\end{equation}
where $|e_{i}\rangle_{12}$ are the factorized basis states%
~(\ref{basis}) and
\begin{mathletters}
\label{g}
\begin{eqnarray}
	\hat{G}_{1}(\varphi)&=&\frac{1}{2}
			\left(\matrix{
		0  & -ie^{i\varphi/2}\sin\frac{\varphi}{2}  \cr
		1 &  e^{i\varphi/2}\cos\frac{\varphi}{2}  \cr} \right)
		\label{g1} \\
\hat{G}_{2}(\varphi)&=&\frac{1}{2}
		\left(\matrix{
		1  & e^{i\varphi/2}\cos\frac{\varphi}{2} \cr
		0 & -ie^{i\varphi/2}\sin\frac{\varphi}{2} \cr} \right)
\label{g2} \\
	\hat{G}_{3}(\varphi)&=&\frac{1}{2}
			\left(\matrix{
		0  & -ie^{i\varphi/2}\sin\frac{\varphi}{2}  \cr
		-1 &  e^{i\varphi/2}\cos\frac{\varphi}{2}  \cr} \right)
\label{g3} \\		
			\hat{G}_{4}(\varphi)&=&\frac{1}{2}
		\left(\matrix{
		-1  & e^{i\varphi/2}\cos\frac{\varphi}{2} \cr
		0 & -ie^{i\varphi/2}\sin\frac{\varphi}{2} \cr} \right)
\label{g4} \;.
\end{eqnarray}
\end{mathletters}
When the photons are detected, Alice sends the results through the 
classical channel to Bob. Bob is left with the photon of mode 3, and 
applies a local unitary transformation $\hat{U}_{i}(\varphi)$ in correspondence
to the $i$-th result of the Bell-state measurement. As a consequence, 
the output state of the teleportation process is
\begin{equation}
	\rho_{out}=\sum_{i=1}^{4}\hat{U}_{i}(\varphi)
	 \hat{G}_{i}(\varphi)|\psi\rangle_{3} \langle \psi |
	 \hat{G}_{i}(\varphi)^{\dagger}\hat{U}_{i}(\varphi)^{\dagger} 
	\label{out}\;.
\end{equation}
Since the output state has to reproduce the unknown input state 
$|\psi\rangle$ as much as possible, it is evident that to optimize the 
local unitary transformations $\hat{U}_{i}(\varphi)$, one should ``invert''  
$\hat{G}_{i}(\varphi)$. The best strategy 
is suggested by the use of the polar decomposition of the matrices 
$\hat{G}_{i}(\varphi)$,
\begin{equation}
\hat{G}_{i}(\varphi)=\hat{T}_{i}(\varphi)\hat{R}_{i}(\varphi)
\label{polar} \;,
\end{equation}
where $\hat{R}_{i}(\varphi)=\sqrt{\hat{G}_{i}(\varphi)^{\dagger}
\hat{G}_{i}(\varphi)}$ is Hermitian and $\hat{T}_{i}(\varphi)$ unitary, 
so that Bob's optimal local unitary transformations will be
\begin{equation}
\hat{U}_{i}(\varphi)=\hat{T}_{i}(\varphi)^{-1}=\hat{R}_{i}(\varphi)
\hat{G}_{i}(\varphi)^{-1}
\label{polar2} \;.
\end{equation}
Using Eqs.~(\ref{g}), (\ref{polar}) and (\ref{polar2}),
one finds the following Bob's optimal unitary transformations
\begin{mathletters}
\label{u}
\begin{eqnarray}
	\hat{U}_{1}(\varphi)&=&
			\left(\matrix{
		-i \cos\frac{\pi+\varphi}{4} & \sin\frac{\pi+\varphi}{4}  \cr
		ie^{-i\varphi/2}\sin\frac{\pi+\varphi}{4} &  
		e^{-i\varphi/2}\cos\frac{\pi+\varphi}{4}  \cr }\right)
\label{u1} \\
 	\hat{U}_{2}(\varphi)&=&
		\left(\matrix{
		\cos\frac{\pi-\varphi}{4} & -i\sin\frac{\pi-\varphi}{4}  \cr
		e^{-i\varphi/2}\sin\frac{\pi-\varphi}{4} &  
		ie^{-i\varphi/2}\cos\frac{\pi-\varphi}{4}  \cr} \right)
\label{u2} \\
	\hat{U}_{3}(\varphi)&=&
			\left(\matrix{
		i \cos\frac{\pi+\varphi}{4} & -\sin\frac{\pi+\varphi}{4}  \cr
		ie^{-i\varphi/2}\sin\frac{\pi+\varphi}{4} &  
		e^{-i\varphi/2}\cos\frac{\pi+\varphi}{4}  \cr} \right)
\label{u3} \\
	\hat{U}_{4}(\varphi)&=&
		\left(\matrix{
		-\cos\frac{\pi-\varphi}{4} & i\sin\frac{\pi-\varphi}{4}  \cr
		e^{-i\varphi/2}\sin\frac{\pi-\varphi}{4} &  
		ie^{-i\varphi/2}\cos\frac{\pi-\varphi}{4}  \cr} \right)
\label{u4} \;,
\end{eqnarray}
\end{mathletters}
which (once the conditional phase shift is known) can be 
easily implemented using appropriate birefringent plates
and polarization rotators.
It can be checked that, in the special case $\varphi = \pi$, the above 
optimized teleportation protocol coincides with the original one
\cite{qt93}, since one has
$\hat{U}_{1}(\pi)=\sigma_{x}$, 
$\hat{U}_{2}(\pi)=1$, $\hat{U}_{3}(\pi)=-i\sigma_{y}$,
and $\hat{U}_{4}(\pi)=-\sigma_{z}$. 

Finally, we have to check that the proposed teleportation protocol,
even though no
longer with $100 \% $ success when $\varphi \neq 
\pi$, always implies the realization of a true quantum teleportation, 
that cannot be achieved with only classical means. This amounts to check
that the average fidelity of the output state is larger than $2/3$
for  $0 < \varphi < 2\pi $.
For pure qubit states, the average fidelity $F_{av}$ is defined as
\begin{equation}
F_{av}=\frac{1}{4\pi}\int d \Omega \langle \psi |\rho_{out}|\psi 
\rangle \;,
\label{fav}
\end{equation}
where the integral is over the Bloch sphere and $|\psi \rangle $
is the generic input state. Using Eqs.~(\ref{out}) and (\ref{polar2})
one has
\begin{equation}
\langle \psi |\rho_{out}|\psi 
\rangle = \sum_{i=1}^{4}\left|\langle \psi|\hat{R}_{i}(\varphi)|\psi 
\rangle \right|^{2}    \;,
\label{fav2}
\end{equation}
so that, using the explicit expressions for $\hat{R}_{i}(\varphi )$ 
that can be obtained from Eqs.~(\ref{g}),
and performing the average over the Bloch sphere, one finally finds
\begin{equation}
F_{av}(\varphi) =\frac{2}{3}+\frac{1}{3}\sin\frac{\varphi}{2}
\label{fav3} \;,
\end{equation}  
which is larger than the
upper classical bound $F_{av}=2/3$ 
\cite{barnum} for $0< \varphi < 2\pi $, as expected.

In conclusion, we have presented a physical implementation for
the quantum teleportation of the polarization state of single 
photons, such as those produced in spontaneous parametric 
down-conversion, based on a crossed-Kerr nonlinearity. In the ideal 
case, the scheme provides a perfect Bell-state discrimination and it 
could be implemented using the giant nonlinearities already 
demonstrated in atomic gases exploiting EIT~\cite{hau,kash}.

{\it Note added in proof.} After submission, we have become aware
of Ref.~\cite{lukin} which shows that a conditional phase shift $\varphi$
close to $\pi$ could be achieved at single photon level if {\em both}
light pulses are subject to EIT and propagate
with slow but equal group velocities. This fact makes us more confident
on the feasibility of the proposed scheme.

\bibliographystyle{unsrt}

\end{multicols}

\end{document}